\pdfoutput=1
\documentclass[prb,twocolumn,aps]{revtex4}
\usepackage{graphicx,graphics,color,epsfig}
\usepackage{bm}
\usepackage{amsmath}
\usepackage{amssymb}
\usepackage{epstopdf}

\makeatletter

\newcommand{\Rmnum}[1]{\expandafter\@slowromancap\romannumeral #1@}
\newcommand*{\rom}[1]{\expandafter\@slowromancap\romannumeral #1@}
\makeatother

\usepackage{color}

\begin{document}
\title{Theoretical investigation of the superconducting pairing symmetry in a bilayer two-orbital model of pressurized La$_3$Ni$_2$O$_7$}

\author{Yi Gao}
\affiliation{Center for Quantum Transport and Thermal Energy Science, Jiangsu Key Lab on Opto-Electronic Technology, School of Physics and Technology, Nanjing Normal University, Nanjing 210023, China}

\begin{abstract}
We investigate the superconducting pairing symmetry in pressurized La$_3$Ni$_2$O$_7$ based on a bilayer two-orbital model. There are two symmetric bands $\alpha$ and $\gamma$, as well as two antisymmetric ones $\beta$ and $\delta$. It is found that the $\gamma$ band induces considerable ferromagnetic spin fluctuation and prefers an odd-frequency, $s$-wave spin triplet pairing state. The addition of the other bands gradually suppresses the ferromagnetic spin fluctuation and enhances the antiferromagnetic ones. The superconducting pairing then evolves from a spin triplet into a $d$-wave spin singlet, and finally into an $s_\pm$-wave one. The competition between the $d$-wave and $s_\pm$-wave pairings relies on whether the ferromagnetic spin fluctuation is suppressed completely or not.

\end{abstract}

\maketitle

\section{introduction}
The recent discovery of superconductivity with $T_c\approx80$ K in pressurized La$_3$Ni$_2$O$_7$ attracted much attention \cite{discovery}. The superconducting phase under high pressure has an orthorhombic structure of $Fmmm$ space group, with the 3$d_{x^2-y^2}$ and 3$d_{z^2}$ orbitals of Ni dominating the energy bands close to the Fermi level. It can be described by a bilayer two-orbital model \cite{bilayermodel} that captures the key ingredients of the material, such as the band structure in the vicinity of the Fermi level, as well as the Fermi surface topology. Due to the mirror symmetry in the high-pressure phase, the energy bands can be categorized into symmetric and antisymmetric ones with respect to exchanging the layer indices. In literatures, they are denoted as $\alpha,\gamma$ (symmetric) and $\beta,\delta$ (antisymmetric) bands, respectively. Among them, only the $\alpha,\beta,\gamma$ bands intersect the Fermi level and form the Fermi surface, while the $\delta$ band is above the Fermi level and is unoccupied. The superconducting pairing symmetry has been investigated extensively based on the bilayer two-orbital model and similar ones. It is found that the $s_{\pm}$-wave pairing symmetry strongly competes with the $d_{x^2-y^2}$-wave and the $d_{xy}$-wave ones \cite{s1,s2,s3,s4,s5,s6,s7,s8,s9,s10,s11,sd1,sd2,sd3,sd4,sd5,sd6,d1,d2,d3}.

In this work, we also study the superconducting pairing symmetry in pressurized La$_3$Ni$_2$O$_7$ based on the bilayer two-orbital model, however, from a different perspective. We select specific bands to calculate the pairing symmetry driven by spin fluctuation, and in this way, we elucidate the role played by different bands in forming the superconducting pairing symmetry, as well as explain the reason why the $s_{\pm}$-wave pairing symmetry competes with the $d_{x^2-y^2}$-wave and the $d_{xy}$-wave ones. We found that, the $\gamma$ band induces considerable ferromagnetic spin fluctuation and tends to form spin triplet pairing, with $s$-wave pairing symmetry and odd frequency dependence. The addition of the other bands gradually suppresses the ferromagnetic spin fluctuation and enhances the antiferromagnetic ones. When these two kinds of spin fluctuations are comparable in strength, a crossover from spin triplet pairing to spin singlet one occurs, and the pairing symmetry tends to be $d$-wave like. When all the four bands are taken into account, the antiferromagnetic spin fluctuation dominates over the ferromagnetic one and the pairing symmetry finally changes to $s_{\pm}$-wave.

\section{method}

We adopt a bilayer two-orbital model of La$_3$Ni$_2$O$_7$ under pressure \cite{bilayermodel}. The two orbitals 3$d_{x^2-y^2}$ and 3$d_{z^2}$ of Ni are denoted as $x$ and $z$, respectively. The tight-binding part of the Hamiltonian can be written as $H_0=\sum_{\mathbf{k}\sigma}\psi_{\mathbf{k}\sigma}^{\dag}M_{\mathbf{k}}\psi_{\mathbf{k}\sigma}$, where $\psi_{\mathbf{k}\sigma}^{\dag}=(c_{\mathbf{k}1x\sigma}^{\dag},c_{\mathbf{k}2x\sigma}^{\dag},c_{\mathbf{k}1z\sigma}^{\dag},c_{\mathbf{k}2z\sigma}^{\dag})$ and
\begin{eqnarray}
\label{h0}
M_{\mathbf{k}}&=&\begin{pmatrix}
T_{\mathbf{k}}^{x}&t_{\bot}^{x}&V_{\mathbf{k}}&V_{\mathbf{k}}^{'}\\
t_{\bot}^{x}&T_{\mathbf{k}}^{x}&V_{\mathbf{k}}^{'}&V_{\mathbf{k}}\\
V_{\mathbf{k}}&V_{\mathbf{k}}^{'}&T_{\mathbf{k}}^{z}&t_{\bot}^{z}\\
V_{\mathbf{k}}^{'}&V_{\mathbf{k}}&t_{\bot}^{z}&T_{\mathbf{k}}^{z}
\end{pmatrix}.
\end{eqnarray}
Here $c_{\mathbf{k}l\alpha\sigma}^{\dag}$ creates a spin $\sigma$ ($\sigma=\uparrow,\downarrow$) electron with momentum $\mathbf{k}$ in the layer $l$ ($l=1,2$) and orbital $\alpha$ ($\alpha=x,z$). In addition,
\begin{eqnarray}
T_{\mathbf{k}}^{x/z}&=&2t_{1}^{x/z}(\cos k_x+\cos k_y)+4t_{2}^{x/z}\cos k_x\cos k_y+\epsilon^{x/z},\nonumber\\
V_{\mathbf{k}}&=&2t_{3}^{xz}(\cos k_x-\cos k_y),\nonumber\\
V_{\mathbf{k}}^{'}&=&2t_{4}^{xz}(\cos k_x-\cos k_y).
\end{eqnarray}

By defining
\begin{eqnarray}
\label{symmetric_antisymmetric}
d_{\mathbf{k}S\alpha\sigma}&=&\frac{1}{\sqrt{2}}(c_{\mathbf{k}1\alpha\sigma}+c_{\mathbf{k}2\alpha\sigma}),\nonumber\\
d_{\mathbf{k}A\alpha\sigma}&=&\frac{1}{\sqrt{2}}(c_{\mathbf{k}1\alpha\sigma}-c_{\mathbf{k}2\alpha\sigma}),
\end{eqnarray}
the tight-binding Hamiltonian can be written in the symmetric and anti-symmetric basis with respect to exchanging the two layer indices as $H_0=\sum_{\mathbf{k}\sigma}\varphi_{\mathbf{k}\sigma}^{\dag}M^{'}_{\mathbf{k}}\varphi_{\mathbf{k}\sigma}$, where $\varphi_{\mathbf{k}\sigma}^{\dag}=(d_{\mathbf{k}Sx\sigma}^{\dag},d_{\mathbf{k}Sz\sigma}^{\dag},d_{\mathbf{k}Ax\sigma}^{\dag},d_{\mathbf{k}Az\sigma}^{\dag})$ and
\begin{eqnarray}
\label{M'k}
M^{'}_{\mathbf{k}}&=&\begin{pmatrix}
T_{\mathbf{k}}^{x}+t_{\bot}^{x}&V_{\mathbf{k}}+V_{\mathbf{k}}^{'}&0&0\\
V_{\mathbf{k}}+V_{\mathbf{k}}^{'}&T_{\mathbf{k}}^{z}+t_{\bot}^{z}&0&0\\
0&0&T_{\mathbf{k}}^{x}-t_{\bot}^{x}&V_{\mathbf{k}}-V_{\mathbf{k}}^{'}\\
0&0&V_{\mathbf{k}}-V_{\mathbf{k}}^{'}&T_{\mathbf{k}}^{z}-t_{\bot}^{z}
\end{pmatrix}.
\end{eqnarray}
$M^{'}_{\mathbf{k}}$ is block-diagonalized and its eigenvalues can be expressed as
\begin{widetext}
\begin{eqnarray}
\label{eigenvalues}
E_{\mathbf{k}}^{1}&=&E_{\mathbf{k}}^{\gamma}=\frac{1}{2}\{T_{\mathbf{k}}^{x}+t_{\bot}^{x}+T_{\mathbf{k}}^{z}+t_{\bot}^{z}-[(T_{\mathbf{k}}^{x}+t_{\bot}^{x}-T_{\mathbf{k}}^{z}-t_{\bot}^{z})^2+4(V_{\mathbf{k}}+V_{\mathbf{k}}^{'})^2]^{\frac{1}{2}}\},\nonumber\\
E_{\mathbf{k}}^{3}&=&E_{\mathbf{k}}^{\alpha}=\frac{1}{2}\{T_{\mathbf{k}}^{x}+t_{\bot}^{x}+T_{\mathbf{k}}^{z}+t_{\bot}^{z}+[(T_{\mathbf{k}}^{x}+t_{\bot}^{x}-T_{\mathbf{k}}^{z}-t_{\bot}^{z})^2+4(V_{\mathbf{k}}+V_{\mathbf{k}}^{'})^2]^{\frac{1}{2}}\},\nonumber\\
E_{\mathbf{k}}^{2}&=&E_{\mathbf{k}}^{\beta}=\frac{1}{2}\{T_{\mathbf{k}}^{x}-t_{\bot}^{x}+T_{\mathbf{k}}^{z}-t_{\bot}^{z}-[(T_{\mathbf{k}}^{x}-t_{\bot}^{x}-T_{\mathbf{k}}^{z}+t_{\bot}^{z})^2+4(V_{\mathbf{k}}-V_{\mathbf{k}}^{'})^2]^{\frac{1}{2}}\},\nonumber\\
E_{\mathbf{k}}^{4}&=&E_{\mathbf{k}}^{\delta}=\frac{1}{2}\{T_{\mathbf{k}}^{x}-t_{\bot}^{x}+T_{\mathbf{k}}^{z}-t_{\bot}^{z}+[(T_{\mathbf{k}}^{x}-t_{\bot}^{x}-T_{\mathbf{k}}^{z}+t_{\bot}^{z})^2+4(V_{\mathbf{k}}-V_{\mathbf{k}}^{'})^2]^{\frac{1}{2}}\}.
\end{eqnarray}
\end{widetext}
Therefore we get two symmetric bands $\alpha$ and $\gamma$, as well as two anti-symmetric ones $\beta$ and $\delta$. The normal Green's function matrix is defined as $G(k)=(ip_nI-M_\mathbf{k})^{-1}$. Here $I$ is the unit matrix and $k=(\mathbf{k},ip_n)$, with $p_n=(2n-1)\pi T$ and $T$ being the temperature. The matrix elements can be calculated as
\begin{eqnarray}
\label{greensfunction}
G^{s_1s_2}(k)&=&\sum_{s_3}\frac{Q_{\mathbf{k}}^{s_1s_3}Q_{\mathbf{k}}^{*s_2s_3}}{ip_n-E_{\mathbf{k}}^{s_3}},
\end{eqnarray}
where $Q_{\mathbf{k}}$ is the unitary matrix that diagonalizes Eq. (\ref{h0}) into Eq. (\ref{eigenvalues}).

The multiorbital Hubbard interaction is written as \cite{scalapino,kubo}
\begin{widetext}
\begin{eqnarray}
\label{honsite}
H_{int}&=&\sum_{i}\sum_{l}\Bigg[U\sum_{\alpha}n_{il\alpha\uparrow}n_{il\alpha\downarrow}
+(U^{'}-\frac{J_H}{2})\sum_{\alpha>\beta}n_{il\alpha}n_{il\beta}
-J_H\sum_{\alpha>\beta}2\mathbf{S}_{il\alpha}\cdot\mathbf{S}_{il\beta}
+J^{'}\sum_{\alpha\neq\beta}c_{il\alpha\uparrow}^{\dag}c_{il\alpha\downarrow}^{\dag}c_{il\beta\downarrow}c_{il\beta\uparrow}\Bigg]\nonumber\\
&=&\sum_{i}\sum_{l}\Bigg[U\sum_{\alpha}n_{il\alpha\uparrow}n_{il\alpha\downarrow}
+U^{'}\sum_{\alpha>\beta}n_{il\alpha}n_{il\beta}
+J_H\sum_{\alpha>\beta}\sum_{\sigma \sigma^{'}}c_{il\alpha\sigma}^{\dag}c_{il\beta\sigma^{'}}^{\dag}c_{il\alpha\sigma^{'}}c_{il\beta\sigma}
+J^{'}\sum_{\alpha\neq\beta}c_{il\alpha\uparrow}^{\dag}c_{il\alpha\downarrow}^{\dag}c_{il\beta\downarrow}c_{il\beta\uparrow}\Bigg],\nonumber\\
\end{eqnarray}
\end{widetext}
where $U$, $U^{'}$, $J_H$ and $J^{'}$ are the strength of intra-orbital Coulomb interaction, inter-orbital Coulomb interaction, Hund's coupling and pair hopping, respectively. Here $i$ is the index of the lattice site in one layer, $l$ denotes the layer and $\alpha,\beta$ are the orbital indices.

The bare susceptibility $\chi_{0}(q)$ is \cite{anticommutation}
\begin{widetext}
\begin{eqnarray}
\label{X0}
\chi_{0}^{s_1s_4,s_2s_3}(q)
&=&\chi_{0}^{l_1\alpha_1l_4\alpha_4,l_2\alpha_2l_3\alpha_3}(q)\nonumber\\
&=&\frac{1}{2N}\int_{0}^{1/T}d\tau e^{i\omega_n\tau}\sum_{\mathbf{k},\mathbf{k}^{'},\sigma,\sigma^{'}}\langle T_{\tau}c_{\mathbf{k}+\mathbf{q}l_1\alpha_1\sigma}^{\dag}(\tau)c_{\mathbf{k}l_4\alpha_4\sigma}(\tau)  c_{\mathbf{k}^{'}-\mathbf{q}l_2\alpha_2\sigma^{'}}^{\dag}(0)c_{\mathbf{k}^{'}l_3\alpha_3\sigma^{'}}(0)\rangle\nonumber\\
&=&-\frac{T}{N}\sum_k G^{s_2s_4}(k+q)G^{s_1s_3}(k).
\end{eqnarray}
\end{widetext}
Here $s_i=(l_i,\alpha_i)$ is the combined layer and orbital index for $i=1,\ldots,4$. $N$ is the number of unit cells and $q=(\mathbf{q},i\omega_n)$, with $\omega_n=2n\pi T$. 

Within the random-phase approximation, the spin and charge susceptibilities can be written as \cite{anticommutation}
\begin{eqnarray}
\label{RPA}
\chi_{s}(q)&=&[I-\chi_0(q)U_s]^{-1}\chi_0(q),\nonumber\\
\chi_{c}(q)&=&[I+\chi_0(q)U_c]^{-1}\chi_0(q),
\end{eqnarray}
where the nonzero matrix elements of $U_s$ and $U_c$ are
\begin{eqnarray}
\label{Us}
U_{s}^{l\alpha l\beta,l\gamma l\delta}&=&\begin{cases}
U&\text{$\alpha=\beta=\gamma=\delta$},\\
U^{'}&\text{$\alpha=\delta\neq\gamma=\beta$},\\
J_H&\text{$\alpha=\beta\neq\gamma=\delta$},\\
J^{'}&\text{$\alpha=\gamma\neq\beta=\delta$},
\end{cases}
\end{eqnarray}
and
\begin{eqnarray}
\label{Uc}
U_{c}^{l\alpha l\beta,l\gamma l\delta}&=&\begin{cases}
U&\text{$\alpha=\beta=\gamma=\delta$},\\
-U^{'}+2J_H&\text{$\alpha=\delta\neq\gamma=\beta$},\\
2U^{'}-J_H&\text{$\alpha=\beta\neq\gamma=\delta$},\\
J^{'}&\text{$\alpha=\gamma\neq\beta=\delta$},
\end{cases}
\end{eqnarray}
with $l$ being the layer index and $\alpha,\beta,\gamma,\delta$ being the orbital indices. For a given $q$, $\chi_0$, $\chi_s$ and $\chi_c$ are all $16\times16$ hermitian matrices. In order to characterize the spin stoner factor, we define $\alpha_s$ as the largest eigenvalue of the matrix $\chi_0(\mathbf{q},i\omega_n=0)U_s$ in the $\mathbf{q}$ space. If $\alpha_s=1$ at a specific $\mathbf{Q}$, then a static long-range magnetic order with a modulation vector $\mathbf{Q}$ will develop. In addition, the momentum structure of the spin fluctuation is manifested as the largest eigenvalue of the matrix $\chi_s(\mathbf{q},i\omega_n=0)$ and is denoted as $\rho_s(\mathbf{q})$.

Close to $T_c$, the linearized Eliashberg equation can be expressed as \cite{anticommutation,eliashberg}
\begin{eqnarray}
\label{Eliashberg}
\lambda \phi^{s_3s_4}(k)&=&-\frac{T}{N}\sum_q\sum_{s_1,s_2,s_5,s_6} G^{s_1s_2}(k-q)G^{s_6s_5}(q-k)\nonumber\\
&&V^{s_6s_4,s_1s_3}(q)\phi^{s_2s_5}(k-q),
\end{eqnarray}
where $\phi(k)$ is the anomalous self energy and the pairing interaction is
\begin{eqnarray}
\label{V}
V(q)&=&
\frac{1}{2}[3U_s\chi_{s}(q)U_s-U_c\chi_{c}(q)U_c+U_s+U_c],\nonumber\\
\end{eqnarray}
for spin singlet pairing and
\begin{eqnarray}
\label{Vt}
V(q)&=&
\frac{1}{2}[-U_s\chi_{s}(q)U_s-U_c\chi_{c}(q)U_c+U_s+U_c],\nonumber\\
\end{eqnarray}
for spin triplet one, respectively.
We solve Eq. (\ref{Eliashberg}) by the power method \cite{power method} to find the largest positive eigenvalue $\lambda$ and the corresponding $\phi(k)$ is the preferred pairing function. In this way, the layer-, momentum-, orbital- and frequency-dependence of $\phi(k)$ can all be solved self-consistently. In the iterative process, due to the anti-commutation relation of the fermions, the initial input $\phi(k)$ should satisfy \cite{anticommutation,oddfrequency}
\begin{eqnarray}
\label{anticommutation}
\phi^{s_1s_2}(k)&=&
\phi^{s_2s_1}(-k),
\end{eqnarray}
for spin singlet pairing and
\begin{eqnarray}
\label{anticommutationt}
\phi^{s_1s_2}(k)&=&
-\phi^{s_2s_1}(-k),
\end{eqnarray}
for spin triplet one, respectively.
After convergence, the anomalous self energy in the band basis $\Delta(k)$ is calculated as
\begin{eqnarray}
\label{Delta_k}
\Delta(k)&=&Q_{\mathbf{k}}^\dag\phi(k)Q_{-\mathbf{k}}^{*}=Q_{\mathbf{k}}^\dag\phi(k)Q_{\mathbf{k}}.
\end{eqnarray}

Throughout this work, the number of unit cells is set to be $N=64\times64$ and the tight-binding parameters are
$(t_{1}^{x},t_{1}^{z},t_{2}^{x},t_{2}^{z},\epsilon^{x},\epsilon^{z},t_{3}^{xz},t_{4}^{xz},t_{\bot}^{x},t_{\bot}^{z})=$(-0.483,-0.110,0.069,-0.017,0.776,0.409,0.239,-0.034,0.005,-0.635), in units of eV \cite{bilayermodel}. The summations over momentum and frequency in Eqs. (\ref{X0}) and (\ref{Eliashberg}) are both done by fast Fourier transformation, where we use $16384$ Matsubara frequencies ($-16383\pi T\leq p_n\leq16383\pi T$ and $-16382\pi T\leq\omega_n\leq16384\pi T$). The temperature is set to be $T=0.007$ eV ($T\approx80$ K). The interaction strength in Eq. (\ref{honsite}) satisfies $U^{'}=U-2J_H$ and we fix $J_H=J^{'}=U/6$.

\section{results and discussion}

\begin{figure}
\includegraphics[width=1\linewidth]{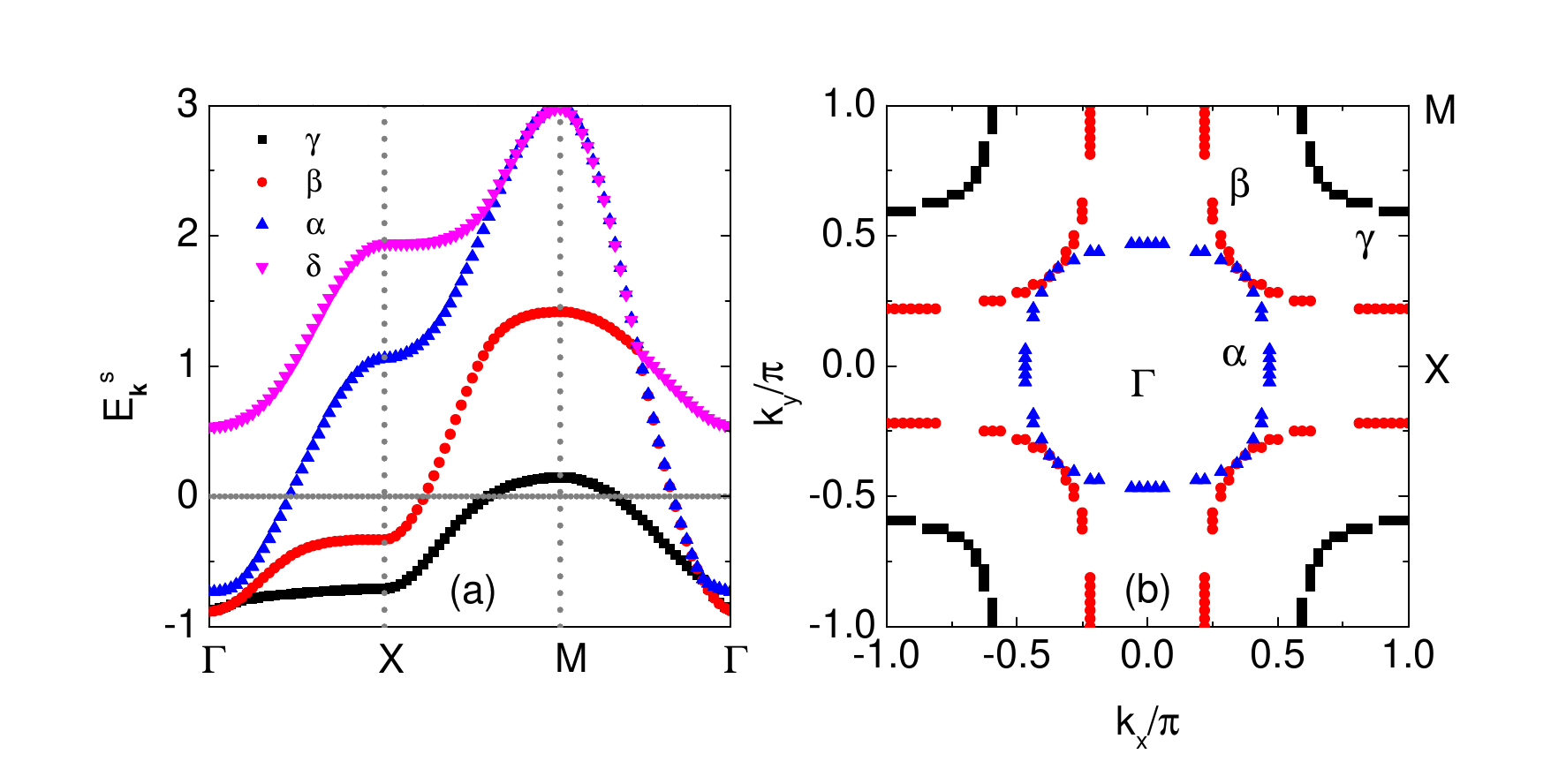}
 \caption{\label{band_structure} (a) The eigenvalues of Eq. (\ref{h0}) along the high-symmetry directions. (b) The corresponding Fermi surface.}
\end{figure}

First we show the band structure and Fermi surface of the bilayer two-orbital model in Figs. \ref{band_structure}(a) and \ref{band_structure}(b), respectively. The four bands are labeled as $\alpha,\beta,\gamma,\delta$, among which only three ($\alpha,\beta,\gamma$) cross the Fermi level and form the Fermi surface, while the $\delta$ band is empty. The filling of each band, defined as $n_s=\frac{2}{N}\sum_\mathbf{k}f(E_{\mathbf{k}}^{s})$ ($s=\alpha,\beta,\gamma,\delta$), is $n_\alpha=0.36$, $n_\beta=0.92$, $n_\gamma=1.72$ and $n_\delta=0$. The filling and Fermi surface shape of the $\beta$ band are similar to those of the hole-doped cuprates, leading to the argument that the $\beta$ band alone will result in the $B_{1g}$ pairing symmetry as in the cuprates \cite{d1}. On the other hand, the filling of the $\alpha$ and $\gamma$ bands is $n_\alpha+n_\gamma=2.08$, close to the optimally electron-doped iron pnictides. In addition, the Fermi surfaces of these two bands are nearly nested by the wave vector $(\pi,\pi)$, a feature also similar to the iron pnictides. Therefore it was suggested that, if only these two bands are considered, the pairing symmetry should be $s_{\pm}$ wave, with a sign change between the $\alpha$ and $\gamma$ Fermi surfaces \cite{d1}. As to the $\delta$ band, since it is unoccupied, it seems that this band will not contribute to the superconducting pairing and can be neglected in the theoretical treatment \cite{d1,sd5}. In the following, we investigate the contribution of these bands to the superconducting pairing, case by case, to verify whether the above arguments are valid or not. Meanwhile, we uncover the peculiarities of La$_3$Ni$_2$O$_7$, in comparison to the cuprates and iron pnictides, despite the many similarities mentioned above.

\begin{figure*}
\includegraphics[width=1\linewidth]{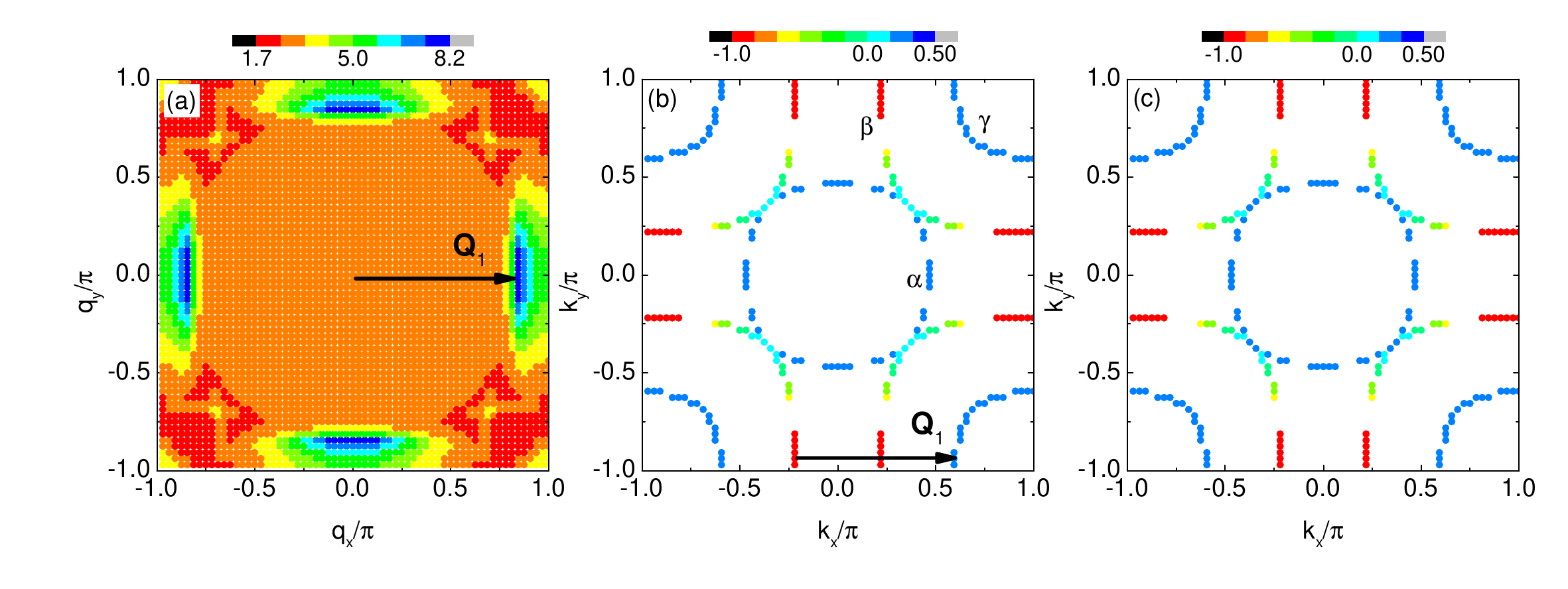}
 \caption{\label{case1} (a) $\rho_s(\mathbf{q})$ when all the four bands are taken into account. (b) The corresponding $\Delta^{ss}(\mathbf{k}_{F},i\pi T)$, with $s=\alpha,\beta,\gamma$. (c) The fitting results of Eqs. (\ref{fit}) and (\ref{fitcase1}). The value of $U$ is 1.16.}
\end{figure*}

Case 1: all the four bands are taken into account. In this case, the spin susceptibility diverges when the interaction strength $U$ is larger than 1.26, therefore we choose the value of $U$ to be 1.16, as in previous studies \cite{s1}, which leads to a spin stoner factor $\alpha_s\approx0.9$, suggesting that the system is close to a magnetic instability, but has not developed a magnetic order. The momentum structure of the spin fluctuation $\rho_s(\mathbf{q})$ is shown in Fig. \ref{case1}(a). It is peaked at $\mathbf{Q}_{1}\approx(\pm0.84\pi,0)$ and $(0,\pm0.84\pi)$ and this wave vector connects the straight portions of the $\beta$ and $\gamma$ Fermi surfaces, with relatively good nesting condition, as shown in Fig. \ref{case1}(b). We then solve Eq. (\ref{Eliashberg}) to get the largest positive eigenvalue $\lambda\approx0.7$ and the corresponding eigenvector $\phi(k)$. The pairing is spin singlet and the pairing function in the band basis is then calculated by Eq. (\ref{Delta_k}). We trace the Fermi momenta $\mathbf{k}_{F}$ of each band and show $\Delta^{ss}(\mathbf{k}_{F},ip_n)$ ($s=\alpha,\beta,\gamma$) at the lowest positive Matsubara frequency ($p_n=\pi T$), which can approximate the pairing function on the Fermi surface. The result is shown in Fig. \ref{case1}(b). We can see, the pairing functions on the $\alpha$ and $\gamma$ Fermi surfaces are of the same sign and are relatively isotropic, while on the $\beta$ Fermi surface, the pairing function is anisotropic and is of the opposite sign with respect to the $\alpha$ and $\gamma$ Fermi surfaces, when the Fermi momenta are connected by the wave vector $\mathbf{Q}_{1}$. The pairing symmetry is $s$-wave, and since there is a sign change between the $\beta$ and $\alpha,\gamma$ Fermi surfaces, the pairing is denoted as $s_\pm$-wave. A further inspection of the $\beta$ band shows that there may exist nodes or gap minima close to the $k_x=\pm k_y$ directions, suggesting a nodal or near nodal behavior in the quasiparticle spectrum. In addition, we can fit the pairing function to
\begin{eqnarray}
\label{fit}
\phi(\mathbf{k},i\pi T)&=&\begin{pmatrix}
f^{x}_{\mathbf{k}}&f^{'x}_{\mathbf{k}}&f^{xz}_{\mathbf{k}}&f^{'xz}_{\mathbf{k}}\\f^{'x}_{\mathbf{k}}&f^{x}_{\mathbf{k}}&f^{'xz}_{\mathbf{k}}&f^{xz}_{\mathbf{k}}\\
f^{xz}_{\mathbf{k}}&f^{'xz}_{\mathbf{k}}&f^{z}_{\mathbf{k}}&f^{'z}_{\mathbf{k}}\\f^{'xz}_{\mathbf{k}}&f^{xz}_{\mathbf{k}}&f^{'z}_{\mathbf{k}}&f^{z}_{\mathbf{k}}
\end{pmatrix},
\end{eqnarray}
with the nonzero matrix elements being
\begin{eqnarray}
\label{fitcase1}
f^{x}_{\mathbf{k}}&=&0.32-0.12(\cos k_x+\cos k_y),\nonumber\\
f^{xz}_{\mathbf{k}}&=&0.2(\cos k_x-\cos k_y),\nonumber\\
f^{z}_{\mathbf{k}}&=&-0.97-0.19(\cos k_x+\cos k_y)+0.28\cos k_x\cos k_y,\nonumber\\
f^{'z}_{\mathbf{k}}&=&1+0.04(\cos k_x+\cos k_y)-0.07\cos k_x\cos k_y.\nonumber\\
\end{eqnarray}
Here, $f^{x}_{\mathbf{k}}/f^{z}_{\mathbf{k}}$ is the intra-orbital and intra-layer pairing function in the $x/z$ orbital, while $f^{'x}_{\mathbf{k}}/f^{'z}_{\mathbf{k}}$ is the intra-orbital but inter-layer one. Furthermore, $f^{xz}_{\mathbf{k}}$ is the inter-orbital and intra-layer pairing function and $f^{'xz}_{\mathbf{k}}$ is the inter-orbital and inter-layer one. The fitting results are shown in Fig. \ref{case1}(c), showing agreement with that in Fig. \ref{case1}(b). Equation (\ref{fitcase1}) suggests the pairing is predominantly on the $z$ orbital. The above pairing symmetry and function agree with previous ones \cite{s1,s7}.

\begin{figure*}
\includegraphics[width=1\linewidth]{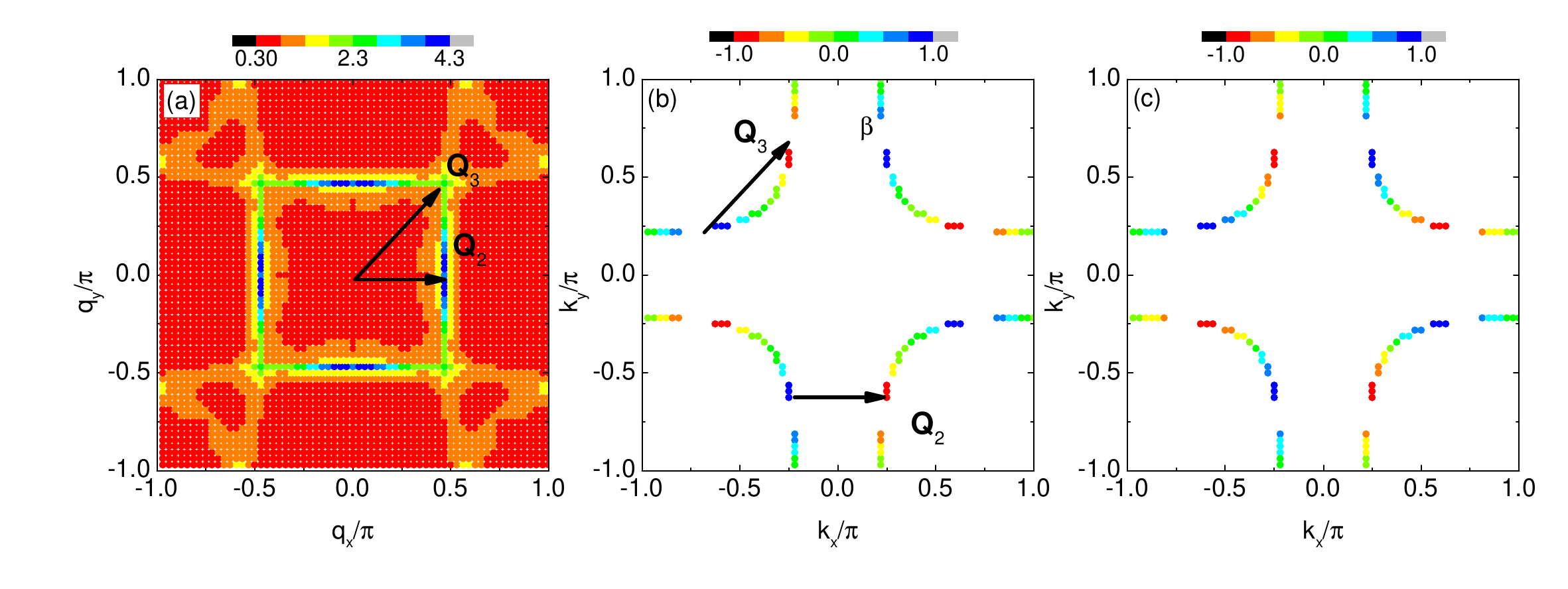}
 \caption{\label{case2} (a) $\rho_s(\mathbf{q})$ when only the $\beta$ band is taken into account. (b) The corresponding $\Delta^{\beta\beta}(\mathbf{k}_{F},i\pi T)$. (c) The fitting results of Eq. (\ref{fitcase2}). The value of $U$ is 4.}
\end{figure*}

Case 2: only the $\beta$ band is considered. In Eqs. (\ref{greensfunction}) and (\ref{Delta_k}), by artificially setting $Q_{\mathbf{k}}^{s_1s_2}=0$ for $s_1=1,\ldots,4$ and $s_2=1,3,4$, we consider only the contribution of the $\beta$ band to the Green's function and in the subsequent calculation. In this case, for $U=1.16$, the calculated $\lambda$ in Eq. (\ref{Eliashberg}) is extremely small ($\lambda<10^{-5}$). The reason is, at this value of $U$, $\alpha_s\approx0.26$, therefore the strength of the spin fluctuation is insufficient to drive superconductivity. For this reason, we set $U=4$ in this case, which leads to $\alpha_s\approx0.9$, comparable to the value in case 1. The spin fluctuation shown in Fig. \ref{case2}(a) is now peaked at $\mathbf{Q}_2\approx(\pm0.5\pi,0)$ and $(0,\pm0.5\pi)$, as well as at $\mathbf{Q}_3\approx(\pm0.5\pi,\pm0.5\pi)$. The $\mathbf{Q}_2$ wave vector connects the straight portions of adjacent $\beta$ Fermi surfaces, while the $\mathbf{Q}_3$ wave vector connects the hot spots within one $\beta$ Fermi surface, as illustrated in Fig. \ref{case2}(b). The largest positive eigenvalue of Eq. (\ref{Eliashberg}) is $\lambda\approx0.46$ and the pairing is spin singlet. The corresponding pairing function projected on the $\beta$ Fermi surface is shown in Fig. \ref{case2}(b). Since the pairing interaction mediated by spin fluctuation for spin singlet pairing is repulsive, therefore on the $\beta$ Fermi surface, the pairing function on those Fermi momenta connected by $\mathbf{Q}_2$ and $\mathbf{Q}_3$ are in opposite sign. However this pairing symmetry is $s$-wave since the pairing function maintains its sign as $\mathbf{k}$ rotates by $\frac{\pi}{2}$. In addition, there exist nodes along the $k_x=\pm k_y$ directions, as well as at $k_x=\pm \pi$ and $k_y=\pm \pi$, suggesting a nodal quasiparticle spectrum. The pairing symmetry is not $d_{xy}$- or $d_{x^2-y^2}$-wave like, because these two pairing symmetries cannot fully utilize the spin fluctuation structure in Fig. \ref{case2}(a). For example, if the pairing symmetry is $d_{xy}$/$d_{x^2-y^2}$-wave, then at the Fermi momenta connected by $\mathbf{Q}_3/\mathbf{Q}_2$, there will be no sign change, leading to a smaller $\lambda$ for these two pairing symmetries. The nonzero matrix elements of Eq. (\ref{fit}) can be fitted to
\begin{eqnarray}
\label{fitcase2}
f^{x}_{\mathbf{k}}&=&-f^{'x}_{\mathbf{k}}\nonumber\\
&=&(\sin 2k_x\sin k_y-\sin k_x\sin 2k_y)\nonumber\\
&+&0.84(\sin 3k_x\sin k_y-\sin k_x\sin 3k_y)\nonumber\\
&-&0.53(\sin 6k_x\sin k_y-\sin k_x\sin 6k_y)\nonumber\\
&-&0.35(\sin 7k_x\sin k_y-\sin k_x\sin 7k_y)\nonumber\\
&+&0.31(\sin 5k_x\sin k_y-\sin k_x\sin 5k_y),\nonumber\\
f^{xz}_{\mathbf{k}}&=&-f^{'xz}_{\mathbf{k}}\nonumber\\
&=&-0.64(\sin 2k_x\sin k_y+\sin k_x\sin 2k_y)\nonumber\\
&+&0.34\sin k_x\sin k_y\nonumber\\
&-&0.32(\sin 3k_x\sin k_y+\sin k_x\sin 3k_y)\nonumber\\
&+&0.31(\sin 6k_x\sin k_y+\sin k_x\sin 6k_y)\nonumber\\
&+&0.25\sin 2k_x\sin 2k_y,\nonumber\\
f^{z}_{\mathbf{k}}&=&-f^{'z}_{\mathbf{k}}\nonumber\\
&=&0.37(\sin 2k_x\sin k_y-\sin k_x\sin 2k_y),\nonumber\\
\end{eqnarray}
with the fitting results shown in Fig. \ref{case2}(c). From Eq. (\ref{fitcase2}) we can see, the pairing on the $x$ orbital is dominant and there is an anti-phase relation between the intra- and inter-layer ones. Clearly we can see, although the Fermi surface and filling of the $\beta$ band resemble those of the hole-doped cuprates, there are distinctions between their spin fluctuation structure. The latter is peaked around $(\pi,\pi)$, leading to the $d_{x^2-y^2}$ pairing symmetry \cite{cuprates}.

\begin{figure*}
\includegraphics[width=1\linewidth]{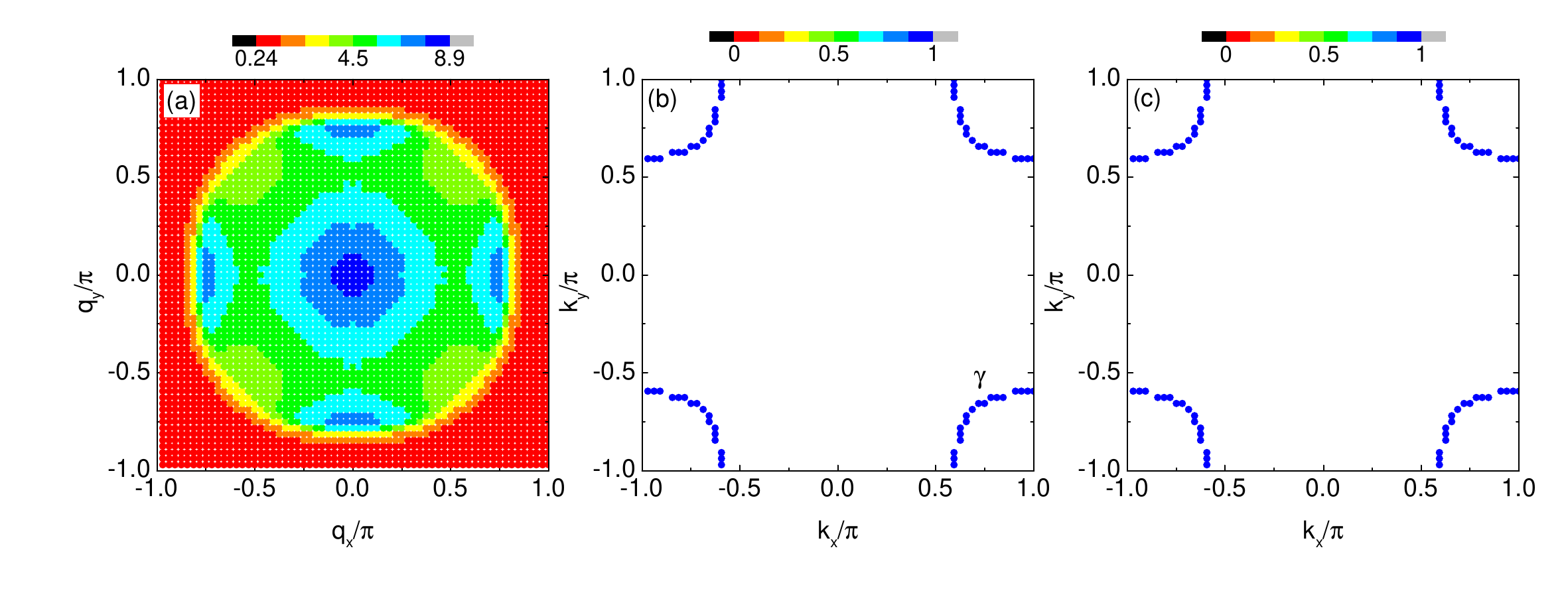}
 \caption{\label{case3} (a) $\rho_s(\mathbf{q})$ when only the $\gamma$ band is taken into account. (b) The corresponding $\Delta^{\gamma\gamma}(\mathbf{k}_{F},i\pi T)$. (c) The fitting results of Eq. (\ref{fitcase3}). The value of $U$ is 2.08.}
\end{figure*}

Case 3: only the $\gamma$ band is considered. From Fig. \ref{band_structure}(a) we can see, the band width of the $\gamma$ band is smaller than the other three, and its Fermi velocity on the Fermi surface is also smaller compared to those on the $\alpha$ and $\beta$ Fermi surfaces. Therefore this band may play a significant role in the formation of superconductivity and we investigate this band alone by setting $Q_{\mathbf{k}}^{s_1s_2}=0$ for $s_1=1,\ldots,4$ and $s_2=2,3,4$ in Eqs. (\ref{greensfunction}) and (\ref{Delta_k}). Similar to case 2, we set $U=2.08$ in this case, leading to $\alpha_s\approx0.9$. The spin fluctuation structure in Fig. \ref{case3}(a) shows a broad peak around $\mathbf{q}=(0,0)$, suggesting a dominant ferromagnetic spin fluctuation. This kind of spin fluctuation usually suppresses the spin singlet pairing and leads to the spin triplet one and indeed we found, the largest positive eigenvalue of Eq. (\ref{Eliashberg}) is $\lambda\approx1.5$ for spin triplet pairing and $\lambda\approx0.8$ for spin singlet one. Thus in this case, spin triplet pairing is preferred. In addition, the pairing function satisfies $\Delta^{\gamma\gamma}(\mathbf{k},ip_n)=-\Delta^{\gamma\gamma}(\mathbf{k},-ip_n)$, i.e., it is odd-frequency pairing \cite{oddfrequency}. Together with the fermionic anti-commutation relation, it further satisfies $\Delta^{\gamma\gamma}(\mathbf{k},ip_n)=\Delta^{\gamma\gamma}(-\mathbf{k},ip_n)$. Thus the superconducting pairing is odd-frequency, even-$\mathbf{k}$ and spin triplet. The pairing function on the $\gamma$ Fermi surface is shown in Fig. \ref{case3}(b), exhibiting an isotropic $s$-wave pairing symmetry. Therefore the quasiparticle spectrum is fully gapped and the pairing function can be fitted to
\begin{eqnarray}
\label{fitcase3}
f^{z}_{\mathbf{k}}&=&f^{'z}_{\mathbf{k}}\nonumber\\
&=&1-0.5(\cos k_x+\cos k_y)\nonumber\\
&+&0.1(\cos 3k_x+\cos 3k_y),
\end{eqnarray}
with the fitting results shown in Fig. \ref{case3}(c). The pairing is on the $z$ orbital and the intra- and inter-layer ones are in the same sign. Since the $\gamma$ band is symmetric and its Fermi surface is dominated by the $z$ orbital \cite{bilayermodel}, the pairing function on it is thus proportional to $f^{z}_{\mathbf{k}}+f^{'z}_{\mathbf{k}}$. We have further verified that, at $U=1.16$, the above conclusions do not change qualitatively, with a smaller values of $\alpha_s\approx0.5$ and $\lambda\approx0.09$.

\begin{figure*}
\includegraphics[width=1\linewidth]{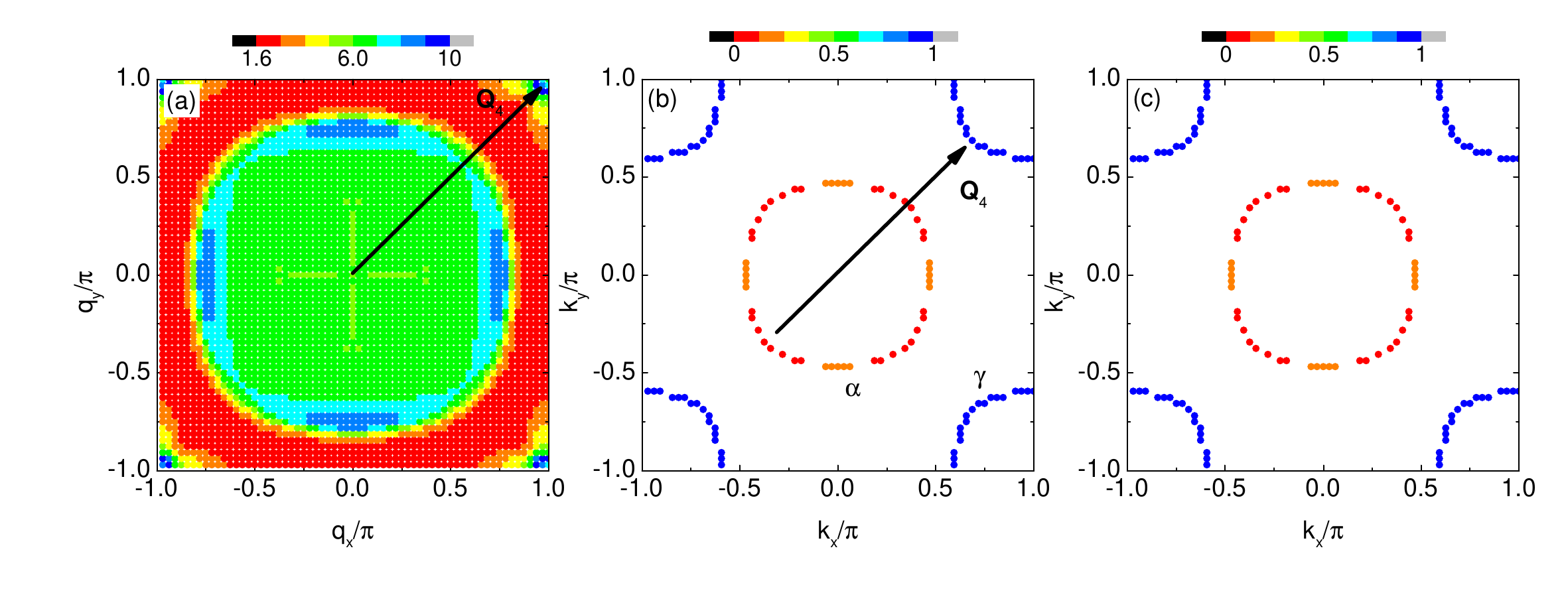}
 \caption{\label{case4} (a) $\rho_s(\mathbf{q})$ when only the $\alpha$ and $\gamma$ bands are taken into account. (b) The corresponding $\Delta^{ss}(\mathbf{k}_{F},i\pi T)$, with $s=\alpha,\gamma$. (c) The fitting results of Eq. (\ref{fitcase4}). The value of $U$ is 1.7.}
\end{figure*}

Case 4: only the $\alpha$ and $\gamma$ bands are considered. By setting $Q_{\mathbf{k}}^{s_1s_2}=0$ for $s_1=1,\ldots,4$ and $s_2=2,4$, we consider only the $\alpha$ and $\gamma$ bands. The reason we consider only these two bands is because, their Fermi surfaces are nested by $(\pi,\pi)$ and their filling is $2.08$, both of which are similar to the optimally electron-doped iron pnictides. Since it is well accepted that the iron pnictides have an $s_{\pm}$ pairing symmetry, we want to know whether it is the same in La$_3$Ni$_2$O$_7$. In this case we choose $U=1.7$ to reach $\alpha_s\approx0.9$. The spin fluctuation structure is shown in Fig. \ref{case4}(a), with a peak at $\mathbf{Q}_4\approx(\pm\pi,\pm\pi)$, as well as a very broad high-intensity region surrounding $\mathbf{q}=(0,0)$. The former is originated from the nesting of the $\alpha$ and $\gamma$ Fermi surfaces and the latter is from the intra-band scattering within the $\gamma$ Fermi surface, as in case 3. The $\mathbf{Q}_4$ spin fluctuation prefers a spin singlet pairing that changes sign between those $\mathbf{k}_{F}$ connected by $\mathbf{Q}_4$, while the $\mathbf{q}=(0,0)$ one prefers a spin triplet pairing. The final pairing symmetry is a competition between these two kinds of spin fluctuations. In this case, the calculated $\lambda\approx1.09$, again in favor of the spin triplet pairing, similar to case 3. Furthermore, since the triplet pairing interaction in Eq. (\ref{Vt}) is attractive, the $\mathbf{Q}_4$ peak in the spin fluctuation further favors a pairing that displays the same sign at those $\mathbf{k}_{F}$ connected by $\mathbf{Q}_4$. The calculated pairing function projected onto the $\alpha$ and $\gamma$ Fermi surfaces is shown in Fig. \ref{case4}(b), it is again odd-frequency, even-$\mathbf{k}$ and spin triplet, as in case 3 and the pairing symmetry is still $s$-wave. The magnitude of the pairing function on the $\gamma$ Fermi surface is larger while that on the $\alpha$ Fermi surface is smaller, and they are in the same sign since they are nested by $\mathbf{Q}_4$. The reason why the pairing symmetry is not $s_{\pm}$-wave singlet is because, the $\gamma$ band is narrow and has a smaller Fermi velocity, while the $\alpha$ band is wider and has a larger Fermi velocity, therefore the intra-band scattering within the $\gamma$ Fermi surface surpasses the inter-band one between the $\alpha$ and $\gamma$ Fermi surfaces, leading to a broad high-intensity region surrounding $\mathbf{q}=(0,0)$ in the spin fluctuation. In contrast, in the iron pnictides, the two Fermi surfaces have comparable Fermi velocities and the inter-band scattering between them is dominant, without the high-intensity region surrounding $\mathbf{q}=(0,0)$. Therefore, it is the Fermi velocity mismatch that leads to a different pairing symmetry as compared to the iron pnictides. The nonzero matrix elements of Eq. (\ref{fit}) can now be fitted to
\begin{eqnarray}
\label{fitcase4}
f^{z}_{\mathbf{k}}&=&f^{'z}_{\mathbf{k}}\nonumber\\
&=&1-0.3(\cos k_x+\cos k_y)\nonumber\\
&+&0.1(\cos 3k_x+\cos 3k_y),
\end{eqnarray}
with the fitting results shown in Fig. \ref{case4}(c). Similar to case 3, the pairing is on the $z$ orbital and is in the same sign between the intra- and inter-layer ones. Since the $\gamma$ Fermi surface is mostly $z$ orbital, while the $\alpha$ Fermi surface is a mixing of the $x$ and $z$ orbitals, therefore the gap magnitude on the $\alpha$ Fermi surface is smaller and on the $\gamma$ Fermi surface, it is $\propto f^{z}_{\mathbf{k}}+f^{'z}_{\mathbf{k}}$. Also we have verified that, at $U=1.16$, the above conclusions do not change qualitatively, with a smaller values of $\alpha_s\approx0.6$ and $\lambda\approx0.14$.

\begin{figure*}
\includegraphics[width=1\linewidth]{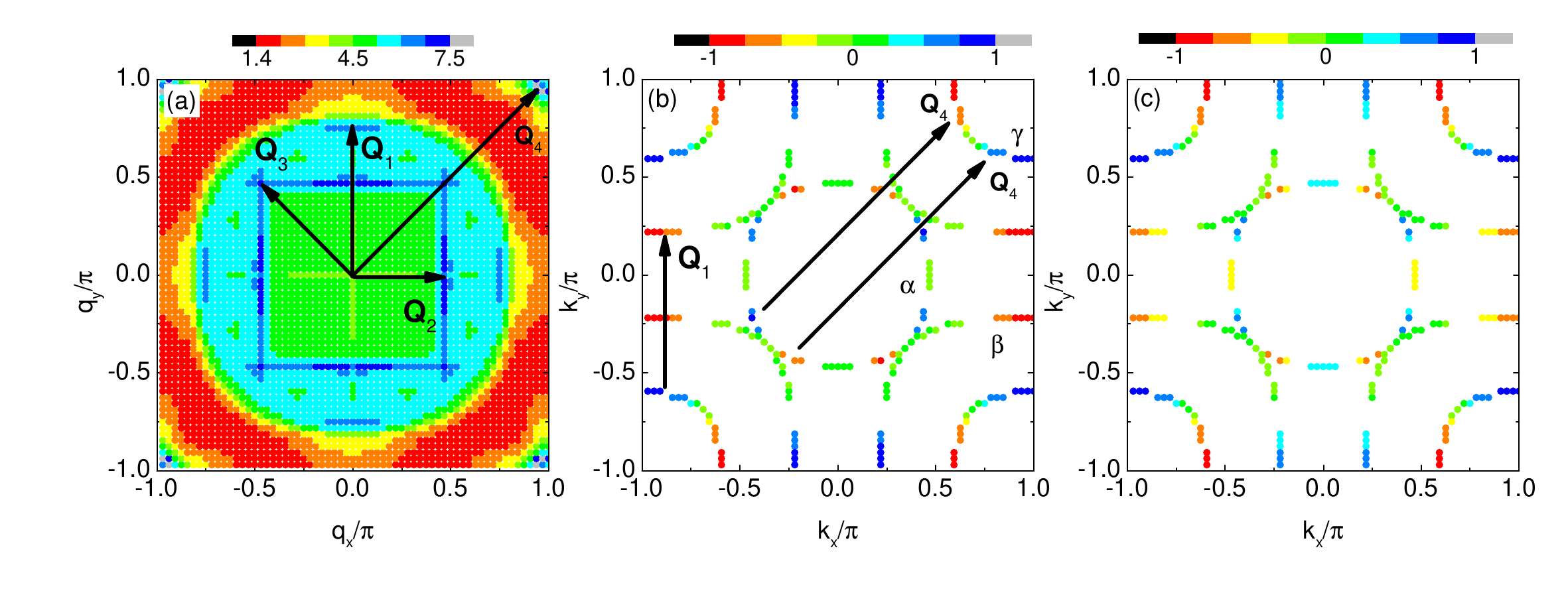}
 \caption{\label{case5} (a) $\rho_s(\mathbf{q})$ when only the $\alpha, \beta$ and $\gamma$ bands are considered. (b) The corresponding $\Delta^{\gamma\gamma}(\mathbf{k}_{F},i\pi T)$, $5\times\Delta^{\alpha\alpha}(\mathbf{k}_{F},i\pi T)$ and $5\times\Delta^{\beta\beta}(\mathbf{k}_{F},i\pi T)$. (c) The fitting results of Eq. (\ref{fitcase5}). The magnitude on the $\alpha$ and $\beta$ Fermi surfaces is multiplied by 5. The value of $U$ is 1.5.}
\end{figure*}

Case 5: only the $\alpha, \beta$ and $\gamma$ bands are considered. We then set $Q_{\mathbf{k}}^{s_1s_2}=0$ for $s_1=1,\ldots,4$ and $s_2=4$ to neglect the contribution of the $\delta$ band. Since the $\delta$ band is unoccupied, we want to verify whether it can be neglected in the theoretical investigation. The value of $U$ is chosen to be 1.5 in this case, which leads to $\alpha_s\approx0.9$. The spin fluctuation structure shown in Fig. \ref{case5}(a) exhibits multiple peaks with comparable intensity, as well as a high-intensity region surrounding $\mathbf{q}=(0,0)$. The $\mathbf{Q}_1$ peak stems from the $\beta-\gamma$ scattering, as shown in Fig. \ref{case1}, while the $\mathbf{Q}_2$ and $\mathbf{Q}_3$ peaks are originated from the $\beta-\beta$ scattering shown in Fig. \ref{case2} and the $\mathbf{Q}_4$ peak is from the $\alpha-\gamma$ scattering shown in Fig. \ref{case4}. All the four peaks in the spin fluctuation prefer a spin singlet pairing. On the other hand, the high-intensity region surrounding $\mathbf{q}=(0,0)$ is from the $\gamma-\gamma$ scattering and it prefers a spin triplet pairing. The competition between them finally leads to a spin singlet pairing with $\lambda\approx0.76$ and the pairing function is shown in Fig. \ref{case5}(b). The pairing symmetry is $d_{x^2-y^2}$-wave (nodal along the $k_x=\pm k_y$ directions), and the gaps on the $\alpha$ and $\beta$ Fermi surfaces are much smaller than that on the $\gamma$ Fermi surface. Clearly, the pairing symmetry is different from Fig. \ref{case1} where all the four bands are considered. The reason is, in the absence of the $\delta$ band, the high-intensity region surrounding $\mathbf{q}=(0,0)$ is still present. In the spin singlet pairing case, it will lead to sign changes with small momentum displacement. Since the $\gamma$ Fermi surface has a smaller Fermi velocity, this ferromagnetic spin fluctuation affects it more effectively by changing the gap sign within the $\gamma$ Fermi surface. On the contrary, if the four bands are all considered, the high-intensity region surrounding $\mathbf{q}=(0,0)$ becomes suppressed and there will be no sign changes with small momentum displacement, therefore in that case, the gap on the $\gamma$ Fermi surface does not change sign. The pairing function can be fitted to
\begin{eqnarray}
\label{fitcase5}
f^{x}_{\mathbf{k}}&=&-0.13(\cos 2k_x-\cos 2k_y)-0.06(\cos k_x-\cos k_y)\nonumber\\
&+&0.04(\cos k_x\cos 2k_y-\cos 2k_x\cos k_y)\nonumber\\
&-&0.03(\cos 3k_x-\cos 3k_y)\nonumber\\
&-&0.03(\cos k_x\cos 5k_y-\cos 5k_x\cos k_y),\nonumber\\
f^{'x}_{\mathbf{k}}&=&0.13(\cos k_x\cos 2k_y-\cos 2k_x\cos k_y)\nonumber\\
&+&0.05(\cos k_x\cos 3k_y-\cos 3k_x\cos k_y)\nonumber\\
&-&0.04(\cos 3k_x-\cos 3k_y)\nonumber\\
&-&0.03(\cos k_x-\cos k_y),\nonumber\\
f^{xz}_{\mathbf{k}}&=&-0.33+0.28(\cos k_x+\cos k_y)\nonumber\\
&+&0.03(\cos 2k_x+\cos 2k_y),\nonumber\\
f^{'xz}_{\mathbf{k}}&=&-0.12-0.14(\cos 2k_x+\cos 2k_y)\nonumber\\
&+&0.25(\cos k_x\cos 2k_y+\cos 2k_x\cos k_y)\nonumber\\
&+&0.07(\cos k_x\cos 3k_y+\cos 3k_x\cos k_y),\nonumber\\
f^{z}_{\mathbf{k}}&=&0.57(\cos 2k_x-\cos 2k_y)+0.48(\cos k_x-\cos k_y)\nonumber\\
&+&0.48(\cos k_x\cos 2k_y-\cos 2k_x\cos k_y)\nonumber\\
&+&0.12(\cos k_x\cos 3k_y-\cos 3k_x\cos k_y)\nonumber\\
&+&0.09(\cos 3k_x-\cos 3k_y),\nonumber\\
f^{'z}_{\mathbf{k}}&=&(\cos 2k_x-\cos 2k_y)\nonumber\\
&+&0.78(\cos k_x\cos 2k_y-\cos 2k_x\cos k_y)\nonumber\\
&+&0.49(\cos k_x-\cos k_y)\nonumber\\
&+&0.26(\cos k_x\cos 3k_y-\cos 3k_x\cos k_y)\nonumber\\
&+&0.11(\cos 3k_x-\cos 3k_y),\nonumber\\
\end{eqnarray}
with the fitting results shown in Fig. \ref{case5}(c). At $U=1.16$, the above conclusions do not change qualitatively, with a smaller values of $\alpha_s\approx0.68$ and $\lambda\approx0.19$. This case suggests, although the $\delta$ band is unoccupied, it still contributes to antiferromagnetic spin fluctuations and helps to change the pairing from spin triplet to spin singlet.

\begin{figure*}
\includegraphics[width=1\linewidth]{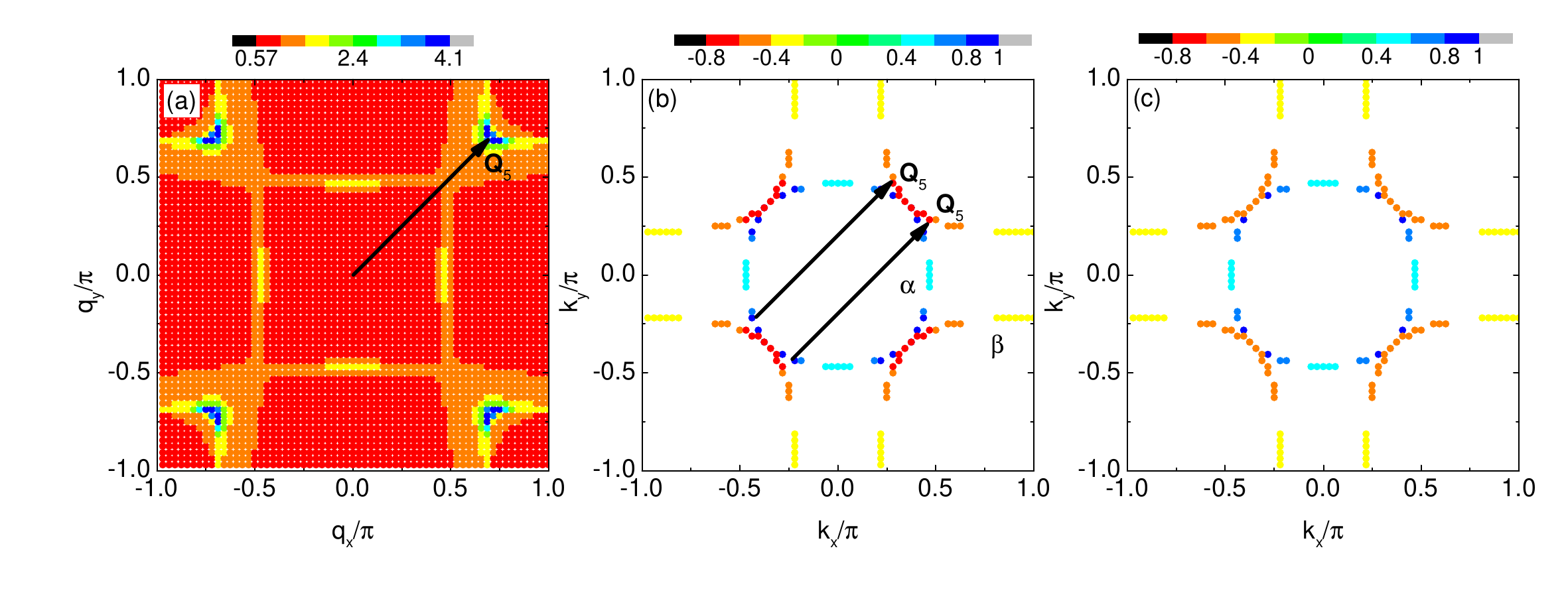}
 \caption{\label{case6} (a) $\rho_s(\mathbf{q})$ when only the $\alpha, \beta$ and $\delta$ bands are considered. (b) The corresponding $\Delta^{ss}(\mathbf{k}_{F},i\pi T)$, with $s=\alpha,\beta$. (c) The fitting results of Eq. (\ref{fitcase6}). The value of $U$ is 2.}
\end{figure*}

Case 6: only the $\alpha, \beta$ and $\delta$ bands are considered. Finally we set $Q_{\mathbf{k}}^{s_1s_2}=0$ for $s_1=1,\ldots,4$ and $s_2=1$ to neglect the contribution of the $\gamma$ band. The value of $U$ is set to be 2 to reach $\alpha_s\approx0.9$. The spin fluctuation structure shown in Fig. \ref{case6}(a) is peaked at $\mathbf{Q}_5\approx(\pm0.7\pi,\pm0.7\pi)$. The largest eigenvalue of Eq. (\ref{Eliashberg}) is $\lambda\approx0.35$ and the pairing is spin singlet. The pairing function on the Fermi surface is shown in Fig. \ref{case6}(b) with the pairing symmetry being similar to that of case 1 and is also $s_{\pm}$-wave, with a sign change between the $\alpha$ and $\beta$ Fermi surfaces. However, the anisotropy on the $\beta$ Fermi surface is smaller, compared to that in case 1. On those Fermi momenta connected by $\mathbf{Q}_5$, the pairing function has a larger magnitude, thus effectively enhancing the value of $\lambda$ of this pairing symmetry. Correspondingly, the quasiparticle spectrum is fully gapped, since there are no nodal or near nodal regions on both the $\alpha$ and $\beta$ Fermi surfaces. The pairing function can be fitted to
\begin{eqnarray}
\label{fitcase6}
f^{x}_{\mathbf{k}}&=&0.21-0.31(\cos k_x\cos 3k_y+\cos 3k_x\cos k_y)\nonumber\\
&-&0.19(\cos 2k_x+\cos 2k_y)\nonumber\\
&+&0.12(\cos k_x\cos 2k_y+\cos 2k_x\cos k_y)\nonumber\\
&+&0.08(\cos 4x\cos 3k_y+\cos 3k_x\cos 4k_y)\nonumber\\
&-&0.07(\cos k_x\cos 6k_y+\cos 6k_x\cos k_y)\nonumber\\
&-&0.07(\cos 7k_x+\cos 7k_y)+0.07(\cos 4k_x+\cos 4k_y)\nonumber\\
&-&0.06(\cos k_x+\cos k_y),\nonumber\\
f^{'x}_{\mathbf{k}}&=&1-0.4(\cos 3k_x+\cos 3k_y)-0.31(\cos k_x+\cos k_y)\nonumber\\
&-&0.3(\cos k_x\cos 2k_y+\cos 2k_x\cos k_y)\nonumber\\
&+&0.12\cos 3k_x\cos 3k_y+0.09(\cos 4k_x+\cos 4k_y)\nonumber\\
&+&0.08(\cos k_x\cos 3k_y+\cos 3k_x\cos k_y)\nonumber\\
&-&0.06(\cos k_x\cos 7k_y+\cos 7k_x\cos k_y)\nonumber\\
&+&0.05(\cos k_x\cos 4k_y+\cos 4k_x\cos k_y),\nonumber\\
f^{xz}_{\mathbf{k}}&=&0.11(\cos 2k_x-\cos 2k_y)\nonumber\\
&+&0.11(\cos 3k_x\cos k_y-\cos k_x\cos 3k_y)\nonumber\\
&+&0.09(\cos k_x-\cos k_y),\nonumber\\
f^{'xz}_{\mathbf{k}}&=&0.19(\cos 3k_x-\cos 3k_y)\nonumber\\
&+&0.08(\cos 2k_x\cos k_y-\cos k_x\cos 2k_y)\nonumber\\
&+&0.07(\cos k_x-\cos k_y),\nonumber\\
f^{z}_{\mathbf{k}}&=&0.06-0.08(\cos 2k_x+\cos 2k_y)\nonumber\\
&-&0.06(\cos k_x+\cos k_y)\nonumber\\
&-&0.05(\cos k_x\cos 3k_y+\cos 3k_x\cos k_y),\nonumber\\
f^{'z}_{\mathbf{k}}&=&-0.37-0.12(\cos k_x+\cos k_y)\nonumber\\
&-&0.08(\cos 3k_x+\cos 3k_y),\nonumber\\
\end{eqnarray}
with the fitting results shown in Fig. \ref{case6}(c), suggesting a dominant inter-layer pairing in the $x$ orbital. In addition, the above results do not change qualitatively at $U=1.16$, with $\alpha_s\approx0.52$ and $\lambda\approx0.03$.

\section{summary}
In summary, we have investigated the superconducting pairing in a bilayer two-orbital model of La$_3$Ni$_2$O$_7$ under pressure. The contribution of each band to the pairing function and symmetry is studied in detail. We found that, the $\gamma$ band itself promotes a ferromagnetic spin fluctuation and prefers a spin triplet pairing. With the addition of the other bands, this ferromagnetic spin fluctuation gradually gets suppressed while other antiferromagnetic ones are enhanced. Correspondingly, the pairing evolves from spin triplet into spin singlet. Although the $\beta$ band and the $\alpha,\gamma$ bands share some similarities with the cuprates and the iron pnictides, respectively, subtle differences in the band structure lead to completely distinct pairing symmetries. Furthermore, although the $\delta$ band is unoccupied, it helps to suppress the ferromagnetic spin fluctuation and to enhance the antiferromagnetic ones through the inter-band scattering and thus cannot be neglected. Due to the small Fermi velocity of the $\gamma$ band, it has more states close to the Fermi level, making it more effective in driving superconductivity. Finally, we address the competition of the $s_\pm$-wave and $d$-wave pairing symmetries reported in literatures. In the present bilayer two-orbital model with the set of tight-binding parameters given by Ref. \onlinecite{bilayermodel}, if all the four bands are taken into account, the spin fluctuation is peaked at an antiferromagnetic wave vector originating from the $\beta-\gamma$ scattering, as shown in Fig. \ref{case1}(a), then the pairing symmetry is most likely $s_\pm$-wave, in agreement with Refs. \onlinecite{s1} and \onlinecite{s7} which use the same tight-binding parameters. In contrast, in other bilayer two-orbital models with different tight-binding parameters \cite{d2,d3}, the ferromagnetic spin fluctuation driven by the intra-band scattering within the $\gamma$ band is still present and the spin fluctuation structure resembles closely to that shown in Fig. \ref{case5}(a) [see Fig. 4(a) in Ref. \onlinecite{d3}], in this case, both our case 5 and Ref. \onlinecite{d3} predicted a $d$-wave pairing symmetry [see Fig. \ref{case5}(b) in our work and Fig. 4(c) in Ref. \onlinecite{d3}]. Therefore, it is the subtle differences in the tight-binding band structure that lead to different pairing symmetries predicted and it relies on whether the ferromagnetic spin fluctuation is suppressed completely or not.

\end{document}